\documentclass{aastex}          
\usepackage{spr-astr-addons}    
\usepackage[colorlinks,anchorcolor=green,citecolor=blue,urlcolor=red,linkcolor=blue]{hyperref}
\usepackage{epsfig, graphicx}
\usepackage{amsmath, amsfonts, amssymb, mathrsfs}
\usepackage{longtable,textcomp, array}
\usepackage[flushleft]{threeparttable}

\begin{document}

\title{Rapid variability of BL Lac 0925+504: interstellar scintillation induced?}

\shorttitle{one-day periodicity in 0925+504}
\shortauthors{Liu et al.}
\maketitle

\author{Jun Liu\altaffilmark{1}}
\email{liuj@xao.ac.cn} 
\and
\author{Xiang Liu\altaffilmark{1,2}}

\altaffiltext{1}{Xinjiang Astronomical Observatory, CAS, 150 Science 1-Street, Urumqi, Xinjiang 830011, P. R. China}
\altaffiltext{2}{Key Laboratory of Radio Astronomy, CAS, 2West Beijing Road, Nanjing, Jiangsu 210008, P. R. China}

\begin{abstract}
Analysis of rapid variability at 4.85 GHz for the BL BLac object 0925+504 is presented and
discussed. The structure functions (SF) are investigated with both refractive and weak interstellar
scintillation (RISS/WISS) models analytically. Parameters obtained with these models are quantitatively compared,
suggesting that the emission region of IDV is remarkably compact and the responsible interstellar scintillation
medium (ISM) lies very close to the observer. Furthermore possible evidence of annual modulation of the
variability timescales is detected in this source. Our findings indicate that the observed rapid variability in 0925+504 is
predominantly caused by a scattering screen located along the line of sight to the source, at a distance of $\sim
110\,pc$ to the observer.

\end{abstract}


\section{Introduction}
\label{sec:intro}

BL Lac objects are known to be highly variable on diverse timescales over all detected wavebands \citep[e.g.][]{Marscher1996,
Fan2000, Chatterjee2008, Gupta2012, Raiteri2013}. In the centimeter regime their variability timescales are often found to be
of the order of a day or less, which is known as intra-day variability (IDV, \citealt{Witzel1986, Heeschen1987}). Evidence
has now accumulated to demonstrate that interstellar scintillation (ISS) in our Galaxy is the dominant cause of such
variability \citep[e.g.][]{Dennett-Thorpe2000, Jauncey2003, Bignall2006, Gabanyi2007, Lovell2008, Marchili2012}.

The ISS theory predicts two regimes of scintillation: (i) strong (diffractive/refractive) scintillation at
low frequencies and (ii) weak scintillation above a transition frequency which is around a few GHz \citep{Walker1998}. While
the diffractive ISS (DISS) is mostly blurred out due to `large size' (comparing
to the pulsar objects) of AGNs, both RISS and WISS are known to be able to induce rapid variability. Evidence for RISS as
being one of the main causes for rapid variability is strongly supported by several compact radio sources (e.g.
0405-385, see \citealp{Jauncey2000}; J1819+385, see \citealp{Dennett-Thorpe2000)} which show large amplitude variations.
Alternatively, the relatively low amplitude variability, as observed in many IDV sources, has been more commonly ascribed to
WISS. Near the transition frequency, either RISS or WISS or a mixture of both are possibly responsible for IDV.

The radio source 0925+504 is a BL Lac object \citep{Plotkin2008} at redshift $z$=0.37 \citep{Healey2008}.
The 5 GHz VLBI image shows a compact core-jet structure \citep{Xu1995}. The direction of radio ejection is very close to the
line of sight with a viewing angle of $3.1^\circ$, which leads to a large Doppler factor of 9.2 \citep{Wu2007}. The source is
associated with very strong $\gamma$-ray emissions \citep{Ackermann2013}. It shows $\sim 13\%$ rms variability at 15 GHz in
the two year OVRO 40m monitoring campaign \citep{Richards2011}. Radio IDV was first discovered at 4.9 GHz in the MASIV
project, where this source showed rapid variability in all the four epochs of MASIV VLA
observations during 2002 and 2003 \citep{Lovell2008}.

We started an IDV survey in the northern sky as early as 2010. BL Lac 0925+504 was one of
the monitoring targets and was observed twice during the project. In 2014, the Urumqi antenna was rebuilt and recovered with
only S/X band to fully serve for the Chinese lunar mission. In order to support the Space VLBI mission RadioAstron and to
investigate the brightness temperature of compact components in AGNs, we proposed two sessions of Effelsberg observations.
The sources were selected according to the RadioAstron observing schedule, and BL Lac 0925+504 was observed in both
sessions.

In this paper, we study the radio variability of BL Lac 0925+504 according to the four epochs of observations.
Both RISS and WISS models are proposed attempting to explain the most likely physical mechanisms behind the observed flux
density variations in this source.

\section{Observation and data reduction}
\label{sec:obs_data}

The IDV observations of 0925+504 were carried out in four epochs, June 19--22 2010 (MJD 55366.1--55369.1),
November 14--19 2012 (MJD 56245.8--56250.2), July 18--21 2014 (MJD 56856.7--56859.3) and September 12--15 2014 (MJD
56912.5--56915.3), in which the first two observing sessions were performed with the Urumqi 25m radio telescope at Nanshan,
while the latter two were with the Max-Planck-Institut f\"ur Radioastronomie (MPIfR) 100m radio telescope at
Effelsberg. At both Urumqi and Effelsberg, the observations of 0925+504 and a number of secondary calibrators were done at a
frequency of 4.85 GHz in cross-scan mode, where the antenna beam pattern was driven repeatedly in azimuth and elevation over
the source position. Frequent switching between targets and calibrators allowed the monitoring of the antenna gain variation
with elevation and time, thus improving the subsequent flux density calibration. Data calibration for both
telescopes were done
in a similar manner, which is well established and enabled high precision (e.g., typical uncertainties for calibrators are
0.4\%$\sim$0.6\% of the measured flux densities, \citealp[see e.g.][]{Liu2012}; see also column 7 of
Table~\ref{tab:var_pars}) flux density measurements  for variability studies.
In short, it consists of the following steps: first Gaussion fitting; then corrections for antenna pointing offsets,
opacity, gain-elevation and gain-time effects; finally scaling the measured antenna temperature to the absolute flux density.
For the detailed data reduction procedures please refer to, e.g. \cite{Kraus2003, Fuhrmann2008}.

\section{Analysis and results}
\label{sec:result}


For a quantitative description of the characteristics of variability, a time series analysis of the light curves was
performed. For each light curve the modulation index $m$, variability amplitude $Y$, reduced $\chi^2$ and intrinsic
modulation index $\overline{m}$, are derived as shown in Table~\ref{tab:var_pars}.
Here we give a brief definition and description of theses quantities, and the reader is referred to e.g.
\citet{Fuhrmann2008, Richards2011} for more details.

The modulation index is related to the standard deviation of the flux density $\Delta_S$ and the mean value of the
flux density $\left< S \right>$ in the time series by
\begin{equation}
 m[\%]=\cfrac{\Delta_S}{\left< S \right>}\cdot 100
\end{equation}
and yields a measure for the strength of the observed variations. The variability amplitude $Y$ is a noise-bias corrected
parameter defined as
\begin{equation}
 Y[\%]=3\sqrt{m^2-m_c^2}
\end{equation}
where $m_c$ is the modulation index of all the observed non-variable calibrators and is a measure of the calibration
accuracy. The intrinsic modulation index $\overline{m}$ \citep{Richards2011} is an alternate estimator to quantify the true
source variability. The definition of $\overline{m}$ involves a two-dimensional maximum-likelihood function
\begin{equation}
\label{eq:likelihood}
\begin{split}
 \mathscr{L}(\overline{m}, S_0) = S_0\left(\prod_{j=1}^N{\frac {1}{\sqrt{ 2\pi \left(
\overline{m}^2{S}_0^2+\sigma_j^2 \right)}}}  \right) \\
\times exp\left[-\frac {1}{2} \sum
_{j=1}^N{
\frac {\left( S_j-S_0 \right)^2}{\overline{m}^2{S}_0^2+\sigma_j^2}}\right]
\end{split}
\end{equation}
where $S_0$ is the true source flux density, $S_j$ the individual flux densities, $\sigma_j$ their errors and $N$ the number
of measurements.
Furthermore, as a criterion to identify the presence of variability, the null-hypothesis of a constant function is examined
via a $\chi^2$-test
\begin{equation}
\chi^2 = \sum _{ j=1 }^{ N }{ { \left( \frac {S_j - \left< S \right>  }{ \sigma_j }
\right)  }^{ 2 } }
\end{equation}
and the reduced value of $\chi^2$
\begin{equation}
\chi_r^2 = \frac { 1 }{ N-1 }\sum _{ j=1 }^{ N }{ { \left( \frac {S_j - \left< S \right>
}{ \sigma_j }
\right)  }^{ 2 } }
\end{equation}
A source is considered to be variable if the $\chi^2$-test gives a probability of $<0.01\%$ for the assumption of constant
flux density (99.99\% significance level for variability).

\begin{table*}[!ht]
  \centering
  \begin{threeparttable}
  \caption{Variability parameters of 0925+504}
  \label{tab:var_pars}
  \begin{tabular}{ccccccccccc} \hline \noalign{\smallskip}
Epoch & Telescope & Obs. & $S_{4.85}$ & $\Delta_S$&
$m$ & $m_c$ & $Y$ & $\chi_r^2$ & $\overline{m}$ & $\tau_0$\\
& & Num. & [Jy] & [Jy] & [\%] & [\%] & [\%] & & [\%] & [day] \\ \hline \noalign{\smallskip}
19.06-22.06.2010 & Urumqi     & 18 & 0.496 & 0.020 & 3.96 & 0.60 & 11.74 & 14.065 & 3.86$^{+1.0}_{-0.6}$  & 0.46$\pm$0.1 \\
14.11-19.11.2012 & Urumqi     & 36 & 0.357 & 0.020 & 5.52 & 0.60 & 16.46 & 8.349  & 5.17$^{+0.9}_{-0.6}$  & 0.64$\pm$0.1 \\
18.07-21.09.2014 & Effelsberg & 15 & 0.336 & 0.013 & 3.96 & 0.50 & 11.78 & 22.392 & 3.87$^{+1.1}_{-0.6}$  & 0.54$\pm$0.1 \\
12.09-15.09.2014 & Effelsberg & 25 & 0.335 & 0.011 & 3.14 & 0.40 & 9.35  & 18.678 & 2.99$^{+0.6}_{-0.4}$  & 1.12$\pm$0.08 \\ \hline
  \end{tabular}
\end{threeparttable}
\end{table*}

To estimate any variability timescales present in our light curves, we employed the structure function (SF) analysis
method (\citealt{Simonetti1985}), which is powerful to deal with unevenly sampled data sets.

\begin{figure}[!ht]
\centering
     \includegraphics[width=0.48\textwidth]{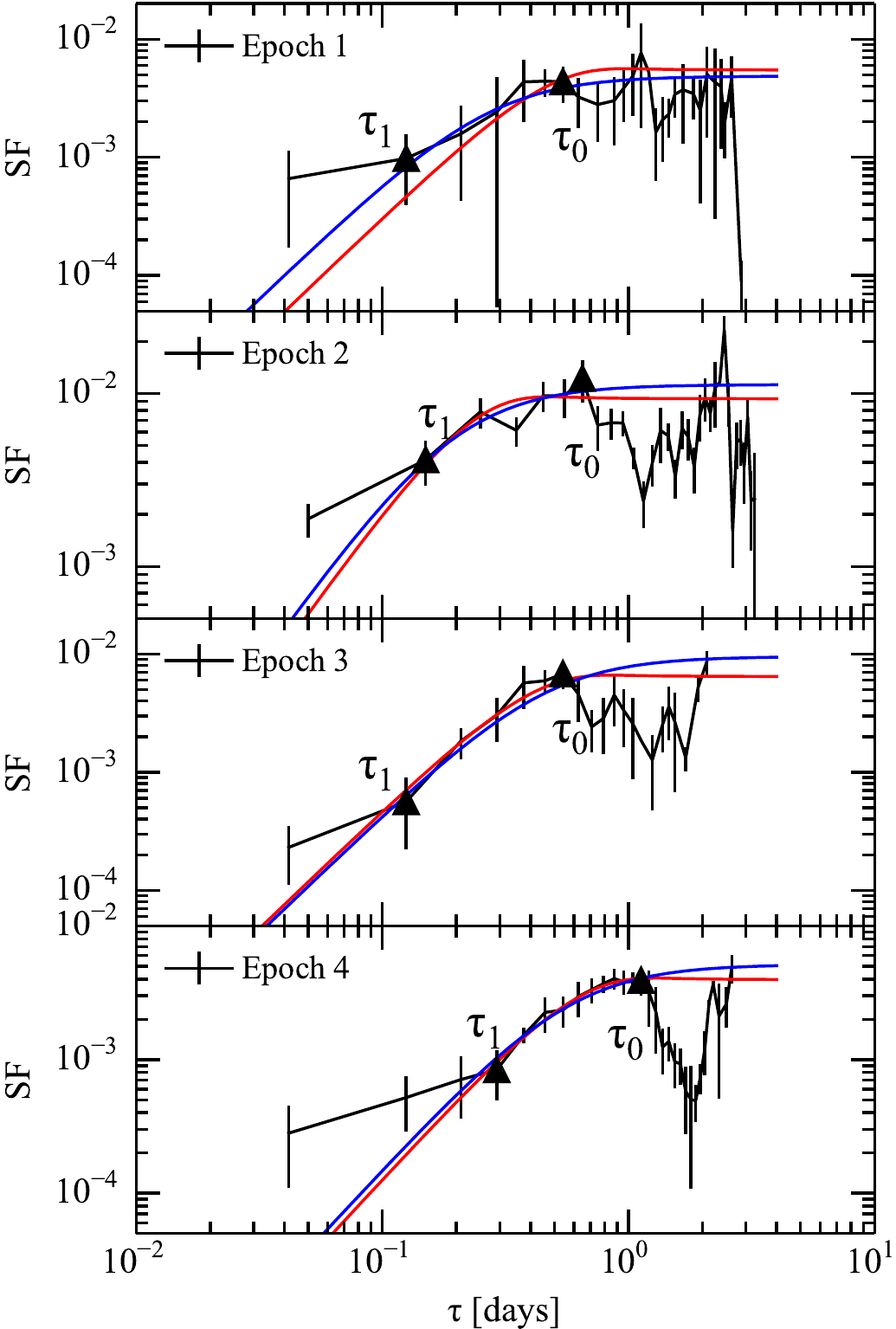}
     \caption{SF analysis for 0925+504. From top to bottom are the results for epoch 1, 2, 3 and 4, respectively. In
each panel the black line indicates the observed SF, while the red/blue line indicates the best RISS/WISS model fit to the
SF curve in the case of a Kolmogorov, thin scattering screen located at a distance of $\sim$ 110 $pc$ to the observer (see
Section \ref{sec:ISS}). The filled triangles at time lags $\tau_1$ and $\tau_0$ represent the time range in which the SF
data points are taken for model fitting (see text).}
      \label{fig:sf}
\end{figure}

As shown in Figure
\ref{fig:sf}, the calculated SF curve (black line in each panel) exhibits manifest features and can typically be
characterized by following behavior: (1) at first it shows a nearly constant or mildly increasing trend at short timescales
($\tau<\tau_1$), which is generally due to noise of the measurements; (2) then it rises up rather steeply when the signal
variability becomes dominant over the noise at $\tau_1\leq\tau\leq\tau_0$. At the time lag $\tau_0$ which corresponds to
the characteristic timescale of the time series, the SF reaches its local maximum; (3) following the rising portion, the
SF saturates and shows either a plateau or a dip for time lags longer than $\tau_0$. Note that the error bars are
generally larger at $\tau>\tau_0$, which should be attributed to the short duration of observation, since at longer
time lags, fewer data points contribute to calculation, making the SF noisy.

Timescale $\tau_0$ is obtained by looking for the local maximum in the SF curve (see Figure \ref{fig:sf}). The associated
uncertainty is given identical to the SF binsize.

In Table \ref{tab:var_pars} we listed the observing information, as well as results of the statistical meaningful
quantities discussed above. In column 1 to 11 are reported the observing epochs, observing facilities, number of effective
measurements, mean flux density, standard deviation of flux density, modulation index, average modulation index of all
calibrators, variability amplitude, reduced $\chi^2$, intrinsic modulation index and characteristic timescales, respectively.

\section{Testing the ISS model}
\label{sec:ISS}

If the fast variability observed in 0925+504 has intrinsic origin, then the size of IDV emitting region should be very small (of the order of light days or less). This will lead to very large apparent brightness temperatures up to $10^{18}$ K,  which is far in excess of the inverse-Compton limit of $\sim 10^{12}$ K \citep{Kellermann1969}. However the observed Doppler factor $D=9.2$ fails to alleviate this problem. Thus the interstellar medium intervening on the line of sight may play a role in the IDV of 0925+504.

In order to produce scintillation responsible for IDV, a few conditions have to be satisfied. Firstly, the distance of the ISM to the observer, which is constrained by the observed range of IDV timescales, should be very small ($10\sim300\,pc$). Secondly, the core of IDV source should be sufficiently compact to induce scintillation. Therefore, study of ISS is a powerful tool not only for the properties of ISM, but also for the compact structure of IDV emission region. In this section, we discuss both RISS and WISS models, and show how the physical properties of ISM and IDV source are constrained and obtained with these models.

\subsection{Refractive Scintillation} \label{subsec:refractive}

In the regime of RISS, the observed variations can be described as the focusing/defocusing action of phase fluctuations in the cones of the diffractive-scale paths, where the measured radiation will be higher/lower as a consequence of the relative speed between the scattering screen and the observer. The analytical model of RISS developed by \citet[and reference therein]{Romani1986} is applied to estimate the RISS predicted SF, which is done by fitting to the observed one. According to \citet{Romani1986}, the auto-correlation function (ACF) for a Kolmogorov, thin scattering screen can be expressed as:
\begin{equation}
\label{eq:acf}
ACF(\tau)=
\frac{Q_0\lambda^4}{(2\pi)^5D^\frac{1}{3}\theta_a^\frac{7}{3}}\,2^\frac{1}{6}\Gamma(\frac{7}{6})M(\frac{7}{6},1,-\frac{
v^2\tau^2 } { 2\theta_a^2D^2} )
\end{equation}
where $Q_0$ is a measure of the scattering strength, $\lambda$ the observing wavelength, $D$ the distance of the scattering screen to the observer, $\theta_a$ the apparent angular size of the source, $M(a,b,x)$ the confluent hypergeometric function, $v$ the transverse velocity of the source relative to the observer.

Furthermore, following the formalism introduced by \citet{Romani1986}, we have
\begin{equation}
\label{eq:theta_i}
 \begin{split}
& \theta_a=(2\pi)^{-\frac{11}{5}}\left[\frac{\Gamma(\frac{7}{6})}{\Gamma(\frac{11}{6})}\frac{9}{5\pi}\right]^{\frac{3}{5}}
 \lambda^\frac{11}{5}Q_0^\frac{3}{5} \\
&  Q_0=3.7\times10^{-18}C_{-4}D\,cm^{-\frac{11}{3}}
  \end{split}
\end{equation}
where $C_{-4}=10^4C_N^2$ defines the turbulence of the scattering screen. Substituting Equation \ref{eq:theta_i} into
Equation \ref{eq:acf}, and expressing $\lambda$ in $meter$, $D$ in $kpc$, $\theta_a$ in $\mu as$, $v$ in $km/s$ and $\tau$ in
$day$, the ACF is derived as
\begin{equation}
 \label{eq:acf1}
 \begin{split}
& ACF(\tau)= 4.64\times10^{-3}\lambda^{-\frac{17}{15}}C_{-4}^{-\frac{2}{5}}D^{-\frac{11}{15}} \\
& \qquad\quad\ \ \times\, M (\frac{7}{6}, 1, -1.67\times10^{-1}\frac{v^2\tau^2}{\theta_a^2 D^2} )
 \end{split}
\end{equation}
Then the theoretical SF is
\begin{equation}
 \label{eq:sf}
 \begin{split}
 SF(\tau)=\ & ACF(0)-ACF(\tau) \\
 =\ & 4.64\times10^{-3}\lambda^{-\frac{17}{15}}C_{-4}^{-\frac{2}{5}}D^{-\frac{11}{15}} \\
 \times & \left[ 1-M (\frac{7}{6}, 1, -1.67\times10^{-1}\frac{v^2\tau^2}{\theta_a^2 D^2} ) \right]
 \end{split}
\end{equation}

We are now able to test the ISS theory by fitting the RISS model to the observed SF curves. The fitted values of the parameters can be used to compare with observations. We start with determining $D$, since physically the value of $D$ should be identical in all epochs. The determination of $D$ is done as follows: a series of fitting is performed to each lightcurve with various initial values ($v=10\sim50\,km/s$, $D=0.01\sim0.3\,kpc$, $C_{-4}=1\sim10000\,m^{-\frac{20}{3}}$). It turns out that $D$ is rather insensitive with the initial conditions. As a consequence, for most cases it converges around a certain value. For each epoch the value is 0.09\,$kpc$, 0.10\,$kpc$, 0.13\,$kpc$, 0.11\,$kpc$, respectively. The average of these values gives $D=0.11\pm0.01\,kpc$. It is worth noting that all the fittings are performed with data cut -- only the data points in the time lag [$\tau_1$, $\tau_0$] are taken into model fitting, since the SF is noisy at both ends, as we addressed in section \ref{sec:result}.

We then fix $D$, and find the best fit of $v$ and $C_{-4}$. Results of $v$, $C_{-4}$ and $\theta_a$ are presented in table \ref{tab:results_riss}. The RISS expected SF curves obtained by model fitting are plotted in red as shown in Figure~\ref{fig:sf}. It is obvious that in all epochs the RISS model fits the data well, indicating that the model we applied is reliable.

\begin{table}[!ht]
\centering
\caption{Values of $v$, $C_{-4}$ and $\theta_a$ obtained by RISS model fitting and subsequent calculations.}
\label{tab:results_riss}
\begin{tabular}{cccc}\hline\noalign{\smallskip}
$v\,[km/s]$ & $C_{-4}$ & $\theta_a\,[\mu as]$ \\ \hline \noalign{\smallskip}
 24.5 &  316.4  &  41.4    \\
 22.5 &   83.8  &  18.6    \\
 21.9 &  210.0  &  32.4    \\
 30.0 &  710.5  &  67.2    \\ \hline
\end{tabular}
\end{table}


\subsection{Weak Scintillation}
In the weak scintillation regime, the flux density variations can be attributed to the effect of weak focusing and defocusing caused by the phase fluctuations which produce a scintillation pattern onto the Earth's orbital plane. The fluctuations are strongest at the Fresnel scale (see \citealp{Narayan1992}), therefore the variability timescale expected for weak scintillation is $\tau_0=r_F/v$, where $r_F=\sqrt{\lambda D/2\pi}$ is the Fresnel scale, $v$ the transverse velocity of the source relative to the observer. In addition, the modulation index is given by \citet{Narayan1992, Walker1998} in the form $m=(\nu_t/\nu)^{17/12}$, where $\nu_t$ is the transition frequency. As demonstrated by \citet{Beckert2002}, the size of AGN bright core component is typically larger than the Fresnel scale, leading to quenched scintillation, where the timescale increases and the modulation index decreases. Then $\tau_0$ and $m$ can be rewritten as \citep{Narayan1992}
\begin{align}
 & \tau_0=\frac{r_F}{v}\left (\frac{\theta_a}{\theta_F} \right) \label{eq:tau0_weak} \\
 & m=\left(\frac{\nu_t}{\nu}\right)^\frac{17}{12} \left( \frac{\theta_F}{\theta_a} \right)^\frac{7}{6} \label{eq:m_weak}
\end{align}
where $\theta_F=r_F/D$ is the Fresnel angular scale, $\theta_a$ the apparent source angular scale. Moreover, an empirical expression of the SF for weak scintillation is given by \citet{Rickett2006}, and written in the following form by
\citet{Lovell2008}:
\begin{equation}
 \label{eq:sf_weak}
 SF(\tau)=2f_c^2m^2\frac{\tau^a}{\tau^a+\tau_0^a}
\end{equation}
where $f_c$ is the fraction of the source flux density in the bright core component, and $1\leq a\leq2$ a constant that depends on the density distribution in the scattering medium. Substituting Equation \ref{eq:tau0_weak} and \ref{eq:m_weak} into Equation \ref{eq:sf_weak}, and expressing $\theta_a$, $\theta_F$ in $\mu as$, $v$ in $km/s$, $\nu$ in $GHz$, $\tau$ in day, and $m$ in percentage, we have
\begin{equation}
 \label{eq:sf_weak1}
 \begin{split}
 SF(\tau)=\ & 4.78\times 10^{-1} f_c^2 \nu_t^\frac{17}{6} D^{-\frac{7}{6}} \theta_a^{-\frac{7}{3}}\\
 \times\ & \frac{\tau^2}{\tau^2+(1.73
D\theta_a/v)^2}
  \end{split}
\end{equation}
where we adopt $a=2$ for a local bubble with low turbulence \citep[e.g.][]{Lovell2008}. We further assume that $\nu_t=4\,GHz$ \citep[see][]{Walker1998} and $f_c=0.4$, then fit Equation \ref{eq:sf_weak1} to the observed SF curves in the same way that we applied for refractive scintillation which is described in section \ref{subsec:refractive}. We again obtained $D=0.11\pm0.01\,kpc$, which is identical to the RISS case. The values of $v$ and $\theta_a$ and $m$ are listed in Tabel \ref{tab:results_weak}, and the WISS expected SF curves are plotted as blue lines in Figure
\ref{fig:sf}.
\begin{table}[!ht]
\centering
\caption{Values of $v$, $\theta_a$ and $m$ for weak scintillation.}
\label{tab:results_weak}
\begin{tabular}{cccc}\hline \noalign{\smallskip}
  $v\,[km/s]$  & $\theta_a\,[\mu as]$ & $m\,[\%]$\\ \hline \noalign{\smallskip}
  36.4 & 52.8 & 12.34  \\
  35.2 & 36.9 & 18.76  \\
  15.5 & 37.9 & 18.19  \\
  16.9 & 51.6 & 12.68  \\ \hline
\end{tabular}
\end{table}

We note that Equation \ref{eq:m_weak} is only strictly valid for $\nu_t\ll\nu$ since it is a asymptotic result \citep{Walker1998}. For $\nu_r\sim\nu$, as shown in Table \ref{tab:results_weak}, the modeled modulation index for each epoch is significantly higher than the observed one, which indicates that $m$ given by Equation \ref{eq:m_weak} is overestimated, and cannot be simply corrected by the `scaling term'. Though degeneracy is found between $f_c$ and $m$ in Equation \ref{eq:sf_weak}, even assuming that the compact core region contains 100\%  of the total flux ($f_c=1$) in the source, the problem still could not be fully alleviated. In addition, the analytical expression of SF developed in \citet{Rickett2006} is empirical, which is determined only by the goodness of fitting of various forms of expressions. As a consequence, the lack of sufficient physical background for Equation \ref{eq:sf_weak} may induce extra uncertainties.

Concerning the validity and uncertainties of the WISS model discussed above, it is unfeasible to verify the competing contributions that RISS and WISS potentially contribute to the observed variability. However, both models imply a source with compact core region $<100\,\mu as$ and predict a local scattering screen with a distance of $\sim\,110\,pc$ on the line of sight, which together lead to scintillation and produce rapid variability.

\subsection{Evidence of Annual Modulation Effect}

\begin{figure}[!ht]
\centering
     \includegraphics[width=0.48\textwidth]{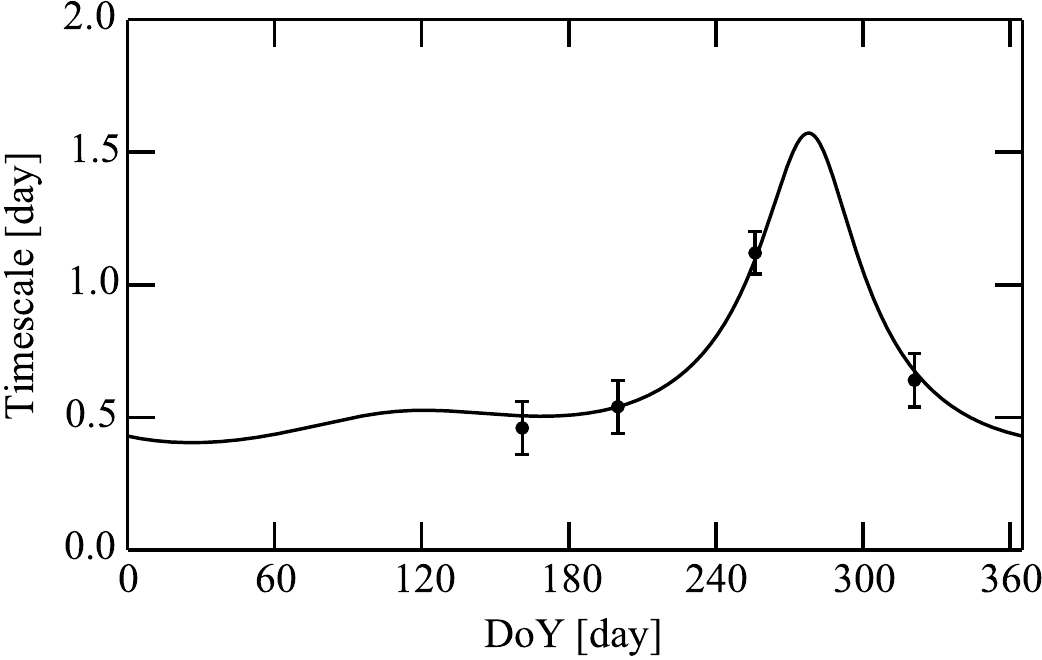}
     \caption{Timescale plotted against day of year. The solid line represents the anisotropic annual modulation model expected timescales, see text for the value of its parameters.}
      \label{fig:annual_modulation}
\end{figure}

We further examine the possible evidence of annual modulation effect \citep[e.g.][]{Dennett-Thorpe2000, Gabanyi2007} in the variability timescales. Due to limited epochs of observations, a highly significant annual modulation fitting to the data cannot be achieved. However, as shown in Figure \ref{fig:annual_modulation}, our result is consistent with the anisotropic annual modulation model with $v_{ra}=2\,km/s$ ,$v_{dec}=8\,km/s$, $D=0.11\,kpc$, $r=7\%$, $\gamma=1^\circ$ ,where $v_{ra}$ and $v_{dec}$ are the screen velocity projected onto right ascension and declination respectively, $D$ the distance of the screen to the observer, $r$ the angular ratio of the anisotropy, and $\gamma$ its position angle. The consistency between the data and the model suggests that a scattering screen intervening on the line of sight may be responsible for the observed variability.

\section{Conclusions}
\label{sec:conclusion}

We modeled the structure function of BL Lac 0925+504 based on the four epochs of observations at 4.85 \,GHz. Both RISS and WISS models are proposed to quantify the physical properties of the scattering medium and the radio source. Our results indicate that the observed rapid variability in 0925+504 is mainly caused by ISS with a distance of $\sim 110\,pc$. Moreover our result is supported by the possible annual modulation effect observed in this source.

Our findings suggest that BL Lac 0925+504 is a promising source for the study of the local ISM as well as AGN physics. More epochs of observations are essential for further testing the annual modulation effect. In the mean time,  multi-wavelength observations will enable us to study the possible wavelength dependent variability, which help to distinguish between different mechanisms of variability and to put new constraints on the source and/or the ISM parameters.

%
\acknowledgments
The authors thank the referee, Dr. Nicola Marchili, for a thorough and insightful review, which improved the quality of the
paper. This paper made use of the data obtained with 25m Urumqi radio telescope of the Xinjiang Astronomical Observatory
(XAO) of the Chinese Academy of Sciences (CAS) and the 100m Effelsberg radio telescope of the Max-Planck-Institut f\"ur
Radioastronomie (MPIfR) in Bonn, Germany. This work was supported by the National Basic Research Program of China (973
program, Grant No. 2015CB857100), the National Natural Science Foundation of China (Grant No. 11273050) and the program of
the Light in China's Western Region (Grant No. YBXM-2014-02).

%
\bibliographystyle{spr-mp-nameyear-cnd}  

%

\end{document}